%% file: main.tex
\documentclass{egpublarxiv}
\usepackage{eg2018}

\ConferenceSubmission   
\electronicVersion 
\PrintedOrElectronic

\title{Joint Material and Illumination Estimation \\
from Photo Sets in the Wild}
\author[Wang et al.]
	{\parbox{\textwidth}{
    \centering 
    Tuanfeng Y. Wang  \qquad
    Tobias Ritschel \quad
    Niloy J. Mitra
    }
    \\
	{
    \parbox{\textwidth}{\centering 
        University College London, UK}
		}
	}

\usepackage{t1enc,dfadobe}

\usepackage[pdftex]{graphicx} \pdfcompresslevel=9

\usepackage{egweblnk}
\usepackage{cite}

\usepackage{amsmath}
\usepackage{paralist}
\usepackage{booktabs}
\usepackage{microtype}
\usepackage{amssymb}
\usepackage{soul}
\usepackage{color}
\usepackage{multirow}
\usepackage{tikz}
\usepackage{tabularx}
\usepackage{units}
\usepackage{mathptmx}
\usepackage{mathrsfs}
\usepackage{wrapfig}

\newcolumntype{C}[1]{>{\centering}m{#1}}

\newcommand{\eg}{e.g.,\ }
\newcommand{\ie}{i.e.,\ }
\newcommand{\etal}{~et~al.\ }

\DeclareGraphicsExtensions{.pdf,.ai,.eps,.jpg}
\def\figurePath{images/}

\def\myfigure#1#2{\begin{figure}[ht]\centering\includegraphics*[width = \linewidth]{\figurePath#1}\caption{#2}\label{fig:#1}\end{figure}}

\def\mycfigure#1#2{\begin{figure*}[t]\centering\includegraphics*[clip, width = \linewidth]{\figurePath#1}\caption{#2}\label{fig:#1}\end{figure*}}

\def\mysection#1#2{\section{#1}\label{sec:#2}}
\def\mysubsection#1#2{\subsection{#1}\label{sec:#2}}
\def\mysubsubsection#1#2{\subsubsection{#1}\label{sec:#2}}

\newcommand{\mypara}[1]{\paragraph*{#1.}}

\newcommand{\refSec}[1]{Section~\ref{sec:#1}}
\newcommand{\refFig}[1]{Figure~\ref{fig:#1}}
\newcommand{\refEq}[1]{Eq.~\ref{eq:#1}}

\newcommand{\synthetic}{\textsc{synthetic}}
\newcommand{\internet}{\textsc{Internet}}
\newcommand{\photos}{\textsc{Photos}}
\newcommand{\docksta}{\textsc{Internet-Docksta}}
\newcommand{\lapd}{\textsc{Internet-lapd}}
\newcommand{\eames}{\textsc{Internet-Eames}}

\definecolor{tobiasColor}{rgb}{0.1,0.6,0.2}

\begin{document}

\teaser{
    \includegraphics[width=\linewidth]{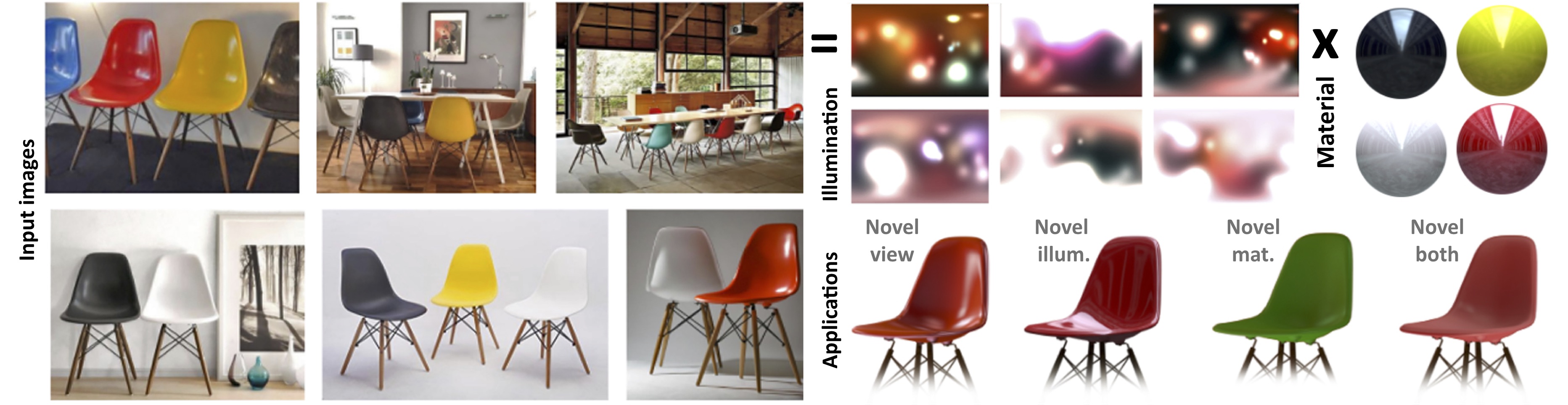}
  \caption{We factor a set of images~{\em (left)} showing objects with different materials (red, yellow, black, white plastic) under different illumination into per-image illumination and per-object material \emph{(top right)} that allows for novel-$x$ applications such as changing view, illumination, material, or mixed illumination/material (red chair in the left-bottom imaged environment) \emph{(bottom right)}.}
  \label{fig:Teaser}
}

\maketitle

\begin{abstract}
Faithful manipulation of shape, material, and illumination in 2D Internet images would greatly benefit from a reliable factorization of appearance into material (\ie diffuse and specular) and illumination (\ie environment maps).
On the one hand, current methods that produce very high fidelity results, typically require controlled settings, expensive devices, or significant manual effort.
To the other hand, methods that are automatic and work on `in the wild' Internet images, often extract only low-frequency lighting or diffuse materials.
In this work, we propose to make use of a set of photographs in order to jointly estimate the non-diffuse materials and sharp lighting in an uncontrolled setting.
Our key observation is that seeing multiple instances of the same material under different illumination (\ie environment), and different materials under the same illumination provide valuable constraints that can be exploited to yield a  high-quality solution  (\ie specular materials and environment illumination)  for all the observed materials and environments.
Similar constraints also arise when observing multiple materials in a single environment, or a single material across multiple environments.
Technically, we enable this by a novel scalable formulation using parametric mixture models that allows for simultaneous estimation of all materials and illumination directly from a set of (uncontrolled) Internet images.
The core of this approach is an optimization procedure that uses two neural networks that are trained on synthetic images to predict good gradients in parametric space given observation of reflected light.
We evaluate our method on a range of synthetic and real examples to generate high-quality estimates, qualitatively compare our results against state-of-the-art alternatives via a user study, and demonstrate photo-consistent image manipulation that is otherwise very challenging to achieve. 
\end{abstract}

\input{introduction}

\input{related_work}
\input{overview}
\input{algorithm}

\input{results}
\input{conclusion}

\bibliographystyle{eg-alpha}
\bibliography{article}
\end{document}

%% file: introduction.tex
\mycfigure{Comparison}{
Comparison to alternatives (projective texturing, average RGB of intrinsic images \protect\cite{Barron2015}).
We see that only a proper separation into specular materials and natural illumination can predict appearance in novel views.
Other approaches miss the highlight, even in the original view (average of intrinsic), or does not move under view changes (projective texturing). Please refer to the accompanying video to judge the importance of moving highlights under view changes. 
}

\mysection{Introduction}{Introduction}

Estimating realistic material (\ie reflectance) and illumination along with object geometry remains a holy grail of shape analysis. While significant advances have been made in the recent years in predicting object geometry and pose from  `in the wild' Internet images, estimation of plausible material and illumination has remained elusive in uncontrolled settings and at a large scale. 

Successful material and illumination estimation, however, will enable unprecedented quality of AR and VR applications like allowing realistic `transfer' of objects across multiple photographs, or inserting high-quality replicas of virtual objects into Internet images. For example, in \refFig{Teaser}, imagine transferring the red chair from one image to another. Currently, this task is challenging as we neither have access to the (red) chair's material, nor the illumination in the target scene.

The naive solution of simply copying and pasting a 2D cutout is unsatisfactory as it easily leads to low fidelity results (\eg unrealistic highlights), and more importantly, does not allow for pose adjustments (see \refFig{Comparison}) or relighting. 

In this paper, we investigate the problem of material and illumination estimation \emph{directly} from `in the wild' Internet images. The key challenge is that material and illumination are never observed independently, but only as the result of the convolving reflection operation with (estimated) normal direction and view direction (assuming access to rough geometry and pose estimates). Thus, in absence of further assumptions, we {\em cannot} uniquely recover material or illumination from single observations (\ie images). 
Instead we reply on {\em linked} observations. 
We observe that often Internet images record the same objects in different environments (\ie  illuminations), or multiple objects in the same environments. 
Such {\em linked} observations among all the materials and illuminations forms a (sparse) {\em observation matrix} providing critical constraints among the observed materials and illumination parameters.
We demonstrate that such a special structure can be utilized to robustly and accurately estimate all the material and illumination parameters through a global optimization. 

We choose a formulation based on the basic rendering equation  in combination with available per-pixel geometry estimation.
However, there are multiple challenges: 
(i)~access to only approximate proxy geometry for the scene objects with rough pose estimates leads to inaccurate normal estimates;
(ii)~partial observations due to view bias (\eg chair backs are photographed less often) and sparsely observed normal directions (\eg flat regions in man-made objects);
 (iii)~working with the rendering equation when updating material and illumination parameters in an inverse problem setup is inefficient in a standard physically-based rendering pipeline; and finally,
(iv)~access to limited data due to sparsely observed joint material-illumination pairs. 

In order to overcome the above challenges, we propose a novel formulation using parametric mixture models. We propose to approximate the reflection operator and its derivative with respect to material and illumination in terms of Anisotropic Spherical Gaussians~(see \cite{xu13aniso}) that can be efficiently utilized to jointly optimize for the materials and illumination at a large scale (\ie involving multiple materials and illuminations). 
This optimization is driven by two neural networks that were trained on a large set of materials and illuminations to predict the gradient the optimization will follow.
For example, in \refFig{Teaser}, we observe 4 different colored (\ie material) chairs under 8 different illuminations (only 6 images shown in the teaser figure) with linked observations. Only using these limited observations, our algorithm extracts high-quality material and illumination estimates, which can then be used for non-trivial image manipulation.

We extensively evaluate our method on both synthetic and real data, both quantitatively and qualitatively (using a user study). We demonstrate that increasing the amount of linked material-illumination observations improves the quality of both the material and illumination estimates. This, in turn, enables novel image manipulations previously considered to be very challenging. 
In summary, our main contributions are: 
(i)~proposing the problem of coupled material and illumination estimation from a set of Internet images; 
(ii)~formulating an efficient and scalable algorithm that allows high-quality material and illumination estimation from a set of images; 
(iii)~using a neural network to approximate the complicated gradient of reflected light with respect to material and illumination parameters; and 
(iv)~utilizing the estimations to enable realistic photo-realistic image manipulations.

%% file: related_work.tex
\mysection{Related Work}{Related Work}
Our goal is to perform an advanced \emph{intrinsic image} decomposition (as a factorization into materials and illumination) using an \emph{image collection} with application in \emph{photo-realistic image manipulation}. %
In the following, we review the related works.

\mypara{Materials and illumination estimation from images}
The classic intrinsic image decomposition problem~\cite{Barrow1978} is highly ambiguous as many shapes, illuminations, and reflectances can explain one observation made.
When geometry and material for the objects in an image are known, finding the illumination is a problem linear in a set of basis images \cite{marschner1997inverse}. 
Reflectance maps~\cite{Horn1989} can also be used to map surface orientation to appearance, allowing for a limited range of applications, such as novel views~\cite{rematas2016views}. 
In absence of such information, alternatives regularize the problem using statistics of each component such as texture~\cite{Shen2008}, or exploit user annotations on Internet images~\cite{Bell2014} to develop a CRF-based decomposition approach, The latter method is widely considered to be the state-of-the-art for uncontrolled observations.

Haber\etal\shortcite{Haber2009} used observations of a single known geometry observed in a small set of images to estimate a linear combination of basis BRDFs and pixel-basis lighting. 
Aittala\etal\shortcite{Aittala2015} capture texture-like materials by fitting SVBRDF using texture statistics to regularize a non-linear optimization on single image capture. 
An alternate recent trend is to use machine learning to solve inverse rendering problems.
Deep learning of convolutional neural networks CNNs~(cf., \cite{Krizhevsky2012}) has been used to decompose Lambertian shading~\cite{Tang2012,Barron2015b}, albedo in combination with other factors~\cite{Narihira2015}, intrinsics from rendered data~\cite{shi2016learning}, decompose images into rendering layers~\cite{innamorati17decomposing}, or multiple materials under the same illumination \cite{georgoulis2016natural}.

A complementary but related problem is shape-from-shading, where again many shapes can explain a given image~\cite{Horn1989}.
Alternately, illumination estimation has made use of shadows on diffuse receivers with known geometry \cite{gibson2001flexible,sato2003illumination,panagopoulos2009robust}.  
Another option is to assume access to outdoor illumination following a parametric sky model~\cite{lalonde2009estimating}.
Approaches to jointly solve for shape, reflectance and illumination in a single image or multiple materials under the same illumination using optimization with priors were suggested \cite{oxholm2015,Lombardi2015}.
Our method, being applicable to direct observation of specular materials under uncontrolled illumination, is more general and works on a large variety of materials in many illuminations. 

Image-based rendering (see monograph~\cite{shum2008image}) can be used to re-create view-dependent appearance when a sufficiently dense set of images is provided.
It then can produce novel views but fails to transfer to novel shapes or novel illuminations.

\mypara{Image and shape collections}
Visual computing has made increasing use of data, particularly image and/or shape collections with the aim to exploit cross observations. 
Starting from illumination \cite{Debevec1998} and its statistics \cite{Dror2001}, measurements of BRDFs~\cite{Matusik2003}, we have seen models of shape~\cite{Ovsjanikov2011exploration}, appearance~\cite{NguyenEG2015}, object pose estimate~\cite{aubry2014seeing}, object texture~\cite{Wang2016TextureTransfer}, object attributes~\cite{HuetingEtAl:Crosslink:2015} made possible by discovering correlation across observations in image and/or 3D model collections. 
In the context of shape analysis, mutual constraints of instances found across images or 3D scenes in the collection have been used to propose room layouts~\cite{craigyu2011furniture},  material assignments~\cite{jain2012material}, or scene color and texture assignments~\cite{Chen2015Decorator}. Instead, we directly estimate materials and illumination, rather than solving an assignment problem. 
%

\mypara{Photo-realistic image manipulation}
While the rendering equation~\cite{Kajiya1986} explains image formation to the largest part with advanced signal processing perspective for the forward theory of light transport~\cite{Ramamoorthi2001}, practical solutions are still lacking for the inverse problem, \ie estimating object materials and illumination directly from (uncontrolled) photographs.
Instead, specialized user interfaces can assist this process \cite{oh2001image}, or multiple images of the same (static) scene can be used to improve relighting and material estimation~\cite{Haber2009}. 
Several manipulations of images are possible, even without knowing the decomposition into shape, illumination and material due to perceptual effects \cite{Khan2006}.

In terms of state-of-the-art image manipulations, 
3-Sweep~\cite{Chen:2013:EEO} propose a generalized cylinder-based interactive tool to efficiently generate part-level models of man-made objects along with inter-part relations, and use them to enable a diverse variety of non-trivial object-level interactions;  
Kholgade\etal\shortcite{OM3D2014} align stock 3D models with input images to estimate illumination and appearance, and use them for impressive object-level image manipulations; while 
Karsch\etal\shortcite{Karsch:TOG:2014} estimates a comprehensive 3D scene model from a single, low dynamic range photograph and uses the information to insert virtual objects into the scene. 

%% file: overview.tex
\mysection{Overview}{Overview}

Starting from a set of linked photographs (\ie multiple objects observed in different shared environments), our goal is to retrieve object geometry with pose predictions and estimate per-object materials and per-environment illuminations. 
The estimated information can then be used to faithfully re-synthesize original appearance and more importantly, obtain plausible view-dependent appearance. 
\refFig{Comparison} shows baseline comparisons to alternative approaches to assign materials to photographed objects. 
We observe that even if the geometry and light is known (we give all the approaches access to our estimated environment maps, if required), the highlights would either be missing (using intrinsic image~\cite{Barron2015} for estimating average albedo), or not move faithfully (\eg with projective texturing) under view changes.

As input, we require a set of photographs of shared objects with their respective masks (see \refFig{Labels}).
In particular, we assume the materials segmentation to be consistent across images.
As output, our algorithm produces a parametric mixture model~(PMM) representation of illumination (that can be converted into a common environment map image) for each photograph and the reflectance parameters for every segmented material.
We proceed in three steps. 

First, we estimate object geometry and pose, and convert all the input images into an unstructured reflectance map for each occurrence of one material in one illumination in \refSec{Aquiring}.
Since we work with very few images collected from the wild, our challenge is that this information is very sparse, incomplete, and often contradict each other.

Second, we solve for illumination for each image and reflectance model parameters for each material in \refSec{Optimization}.
This requires combining a very large number of degrees of freedom, as fine directional lighting details as well as accurate material parameters to be estimated.
The challenge is that a direct optimization can easily involve many variables non-linearly coupled and lead to a cost function that is highly expensive even to evaluate as it involves solving the forward rendering equation, \eg \cite{Lombardi2015}.
For example, representing images and illumination in the pixel basis leads to an order of $10^4$-$10^5$ variables (\eg $128\times 256\times$number-of-environment-maps).
At the same time, evaluating the cost function for every observation pixel would amount to gathering illumination by iterating all pixels in the environment map, \ie an inner loop over all $128\times 256$ environment map pixels inside an outer loop across all the $640\times 480\times$number-of-images-in-the-collection observations. This quickly becomes computationally intractable. 

Instead, we introduce a solution based on parametric mixture-model (PMM) representation of illumination to {\em inverse} rendering, which has been successfully applied to forward rendering~\cite{green2006view,wang2009all,wu2011sparse,xu13aniso,vorba2014line}.
Our core contribution is to take PMM a step further by introducing the parametric mixture reflection operator and an approximation of its gradient, allowing to solve the optimization in a scalable fashion involving many materials and environments.
The gradient approximation uses a neural network to map from observed reflected light, light and material parameters to changes of light and material parameters.
It is trained on a set of synthetic images rendered from many illuminations and many materials.

Third, the estimated material and illumination information can directly be used in standard renderers.
The challenge in such applications is to capture view-dependent effects such as moving highlights.
In \refSec{Applications}, we show applications to manipulating images, changing the illumination and/or material and/or view, transferring materials to objects in other images, or inserting objects into new illumination (see also supplementary materials).

%% file: algorithm.tex
\mysection{Algorithm}{Our Approach}

We now explain our approach in  details. 

\mysubsection{Acquiring Geometry and Reflectance Maps}{Aquiring}

We start from a set of images with the relevant materials segmented consistently across the image collection. Designer websites (\eg Houzz) and product catalogs (\eg Ikea) regularly provide such links. Here we assume that the links are explicitly available as input. First, we establish a mapping between illumination material-pairs and observed appearance.

\myfigure{Labels}{RGB, normal, and segmentation of a typical input image.}

\mypara{Per-pixel labels}
%
For the input images, we used per-pixel orientation (screen-space normals) (\refFig{Labels}) obtained using render-for-CNN~\cite{Su_2015_ICCV} trained on the ShapeNet to retrieve object geometry and pose estimates. We found this to provide better quality normal predictions than those obtained via per-pixel depth~\cite{Eigen2015} and normal~\cite{Wang2015} estimation.

\mypara{Reflectance maps}
The rendering equation \cite{Kajiya1986} states that
\begin{align}
\small
L_\mathrm o(\mathbf x,\mathbf n,\omega_\mathrm o)
=
\underbrace{
L_\mathrm e(\mathbf x,\omega_\mathrm o)
}_\text{Emit}
+
\int_\Omega
\underbrace{
f_\mathrm r(\mathbf x, \omega_\mathrm i, \omega_\mathrm o)
}_\text{BRDF}
\underbrace{
L_\mathrm i(\mathbf x,\omega_\mathrm i)
}_\text{Incom.}
\underbrace{
<\mathbf n, \omega_i>^+
}_\text{Geometry}
\mathrm d\,\omega_\mathrm i,
\label{eq:rend} 
\end{align}
where 
$\mathbf x$ is the position,
$\mathbf n$ the surface normal at location $\mathbf x$,
$\omega_\mathrm o$ the observer direction,
$L_\mathrm o$ is the observed radiance,
$L_\mathrm e$ is light emission,
$L_\mathrm i$ is the incoming illumination, and 
$f_\mathrm r$ the bi-directional reflectance distribution function~(BRDF)~\cite{Nicodemus1965}.

We assume a simplified image formation model that allows for using a slightly generalized variant of reflectance maps~\cite{Horn1989}:
(i)~distant illumination,
(ii)~convex objects, \ie no shadows or inter-reflections, 
(iii)~{\em spatially invariant} BRDFs, and 
(iv)~no emission.
Note that we do {\em not} assume a distant viewer as typical reflectance map does.
This simplifies \refEq{rend} to 
\begin{align}
L_\mathrm o(\omega_\mathrm o, \mathbf n)=
\int_\Omega
f_\mathrm r(\omega_\mathrm i, \omega_\mathrm o)
L_\mathrm i(\omega_\mathrm i)
<\mathbf n, \omega_i>^+
\mathrm d\,\omega_\mathrm i.
\end{align}
A classic reflectance map is parameterized either by normal $\mathbf n$ or by the observer direction $\omega_\mathrm o$.
Instead of making such a split, we take a less structured approach tailored to our problem: an \emph{unstructured reflectance map} (URM) denoted by $\mathcal O$ that uses a list that holds in each entry a tuple of 
(i)~normal $o_\mathrm n$, 
(ii)~half-angle vector $\mathbf{h}$, 
(iii)~observed radiance $o_L$ (cf.\ \refFig{ReflectanceMaps}), and 
(iv)~indices $o_\mathrm m$ and $o_\mathrm i$ of the material and illumination, respectively.
We denote $\mathbf{h}$ as the half-angle vector for front ($-z$) and observer direction, $\mathbf h := (<2\mathbf n,o_\omega>\cdot\mathbf n - o_\omega+(0,0,-1))/2$.
This parametrization will provide a more convenient way to index information.
An example visualization of the URM by projecting the $\mathbf n$ as well as the $\mathbf h$ coordinate using latitude-longitude is seen in \refFig{ReflectanceMaps}.

\myfigure{ReflectanceMaps}{
Schema and actual Unstructured Reflectance Maps of the chairs in the first column of \protect\refFig{Teaser}.
Each point is an observed color for a specific surface orientation $\mathbf n$ and half-angle vector $\mathbf h$.}

\mycfigure{Concept}{
The three main ideas to enable large-scale optimization:
{\textbf (a)} approximating illumination as parametric mixture models and the BRDF as a sum of a diffuse and a specular component; 
{\textbf (b, c)} expressing reflection as a sum of diffuse and specular reflections of individual lobes; and 
{\textbf (d)} approximating derivative of diffuse and specular reflection of ISGs using corresponding neural nets.
}

To acquire the URM from an image with given per-pixel position and orientation, we apply inverse gamma correction such that $o_L$ is in physically linear units.
Note, that although we do not know the absolute scale inside each photo, we do not need it for most applications.
Further, we do not differentiate between objects and consider only their materials (\ie an object with two material parts are essentially treated as two materials).

\mysubsection{Representation}{Representation}

\mypara{Illumination}\label{IlluminationRepresentation}
We use Parametric Mixture Models~(PMMs) to represent illumination.
PMMs have been used for pre-computed light transport \cite{green2006view,tsai2006all,wang2009all}, BTF compression~\cite{wu2011sparse}, interactive rendering~\cite{tokuyoshi2015virtual}, importance sampling~\cite{vorba2014line},  or even in caustic design~\cite{papas2011goal}.
%
A PMM encoded as 
\begin{align}
g(\omega |\mathsf\Theta) := 
\sum_{l=1}^{n_\mathrm p} p(\omega | \Theta_l) 
\approx
L(\omega)
\end{align}
is a sum of $n_\mathrm p$ \emph{lobe} functions $p(\omega|\Theta_l)$ that depend on a parameter vector $\Theta_l$ to approximate, in our setting, the incoming or outgoing light function $L(\omega)$.
All parameter vectors $\Theta_l$ of one PMM are combined in a parameter matrix $\mathsf\Theta$. 
In our case, the domain of $g$ is the sphere $\Omega$ parameterized using  latitude-longitude representation $\omega=(\theta,\phi)\in[0,2\pi)\times[0,\pi)$.

As mode functions, we employ Isotropic Spherical Gaussians~(ISGs) \cite{green2006view,tsai2006all,vorba2014line}.
An ISG lobe has the form $$
p(\omega | \Theta) :=
w
\cdot
\exp(
-\sigma(\omega-\mathbf z)^2
),
$$
where 
$w\in\mathbb R^+$ is the weight of the lobe, 
$\sigma$ is its variance and
$\mathbf z$ the \emph{mean} direction.
Consequently, a lobe is described by parameter vector
$
\Theta=
(
w,
\sigma,
\mathbf z
)
$.
To work with RGB values all weight components $\mathbf w$ in this paper are vector-valued, but the variance parameter $\sigma$ is scalar.
For each image, we use an ISG PMMs with $n_\mathbf p=32$ components to represent unknown illuminations.

\mypara{Material}\label{MaterialRepresentation} 
We assume the material to be of the form
\begin{align}
f_\mathrm r(\omega_\mathrm i,\omega_\mathrm o | \rho)
=
\underbrace{
k_\mathrm d
f_\mathrm d(\omega_\mathrm i,\omega_\mathrm o)
}_\text{Diffuse}
+
\underbrace{
k_\mathrm s
f_\mathrm s(\omega_\mathrm i,\omega_\mathrm o | r)
}_\text{Specular},
\end{align}
a parametric model that can be split into the weighted sum of a diffuse and a specular component $f_\mathrm d$ and $f_\mathrm s$ with weights $k_\mathrm d$ and $k_\mathrm s$, repectively. 
We choose Lambertian as the diffuse model and GGX~\cite{Walter:2007:MMR} that has a single roughness parameter $r$ as the specular model.
The material parameters are therefore a tuple $\rho=(k_\mathrm d,k_\mathrm s,r) \in \mathbb{R}^7$ of RGB diffuse and specular reflectance and a scalar roughness parameter.
We denote the BRDF parameter vector of material $j$ as $\rho^{(j)}$.
Note that we do not need to represent $f_\mathrm r$ using a PMM, which would introduce unnecessary approximation error.

\mysubsection{Reflection}{Reflection}
Using standard notation~\shortcite{arvo1994framework} for light transport, we express reflection as an operator $\mathbf R$, mapping the function of incoming light $L_\mathrm i$ to a function of reflected outgoing light $L_\mathrm o$ :
\begin{align}
L_\mathrm o(\omega_\mathrm o)
=
\mathbf R
(L_\mathrm i
|
\rho
)
(\omega_\mathrm o)
=
\int_\Omega
\underbrace{
L_\mathrm i(\omega_\mathrm i)
}_\text{Illumination}
\underbrace{
f_\mathrm r(\omega_\mathrm i,\omega_\mathrm o|\rho)
}
_\text{BRDF}
\mathrm d\omega_\mathrm i.
\end{align}
When using an ISG to represent the illumination, we suggest to use a \emph{parametric reflection operator} $\mathbf R(\Theta|\rho)$ that maps from a single illumination ISG lobe $\Theta$ and a material $\rho$ to a reflected light.
As we assume the BRDF to be a sum of a diffuse and a specular part, we can similarly define $\mathbf D$ and  $\mathbf S$ that are respectively the diffuse  and the specular-only reflection and $\mathbf R=\mathbf D+\mathbf S$. So, finally we have
\begin{align}
L_\mathrm o(\omega_\mathrm o)
=
\sum_{l=1}^{n_\mathrm l}
\mathbf D(\Theta_l|\rho)+
\mathbf S(\Theta_l|\rho).
\end{align}

\mycfigure{NNEvaluation}{
Evaluation of the neural network.
The first row shows GT renderings with a GT envmap.
The second row shows again GT rendering, but using the GMM fit to the envmap.
This is an upper bound on the NN quality, as it works on the GMM representation.
The third row shows the NN result.
In the horizontal direction, specular results of increasing roughness are followed by the diffuse result in the rightmost column.
The plots on the right below show the error distribution as a function of different parameters.
}

\mysubsection{Formulation}{Optimization}
Our task is to find a set of illuminations and a set of materials that explain the acquired observations (see the previous section).
Next, we describe how to represent reflectance and illumination as well as introduce the parametric reflectance operator, its derivative with respect to material and illumination, and an approximation method for efficient joint optimization for material and illumination given the observations (see 
\refFig{Concept}). 

\mypara{Cost function}
Our main objective function quantifies how well a set of materials and illuminations explain the input observation.
It should be fast to evaluate and allow for an effective computation of its gradient with respect to illuminations and materials in order to be useful in an optimization. We formulate the objective as: 
\begin{align}
c(
\mathsf\Theta,
\rho|
\mathcal O
) :=
\underbrace{
    \sum_{\mathbf o\in\mathcal O}
    \left\|
    o_\mathrm L
    -
    \sum_{l=1}^{n_\mathrm p}
    \mathbf R\left(
    \Theta_l^{(o_\mathrm i)}|\rho^{(o_\mathrm m)}\right)(o_\omega
    )
    \right\|^2
}_\text{Data}
+
\underbrace
{
    \vphantom{
        \sum_l^{n_\mathrm p}
        \left|\right|_2^2
    }
    \lambda
    p(\mathsf\Theta)
}_\text{Prior}
.
\end{align}
The gradient of this function with respect to the illumination and material comprises of evaluating $\mathbf R$, which involves convolving an illumination lobe with the BRDF.
This is both costly to compute and we need to find its derivative.
To this end, we will employ a learning-based approach, as described next.

\mypara{Neural network}
The input to this neural network~(NN) is the parameters of a single illumination lobe, the material parameters, and the observation direction $\omega$.
We call this approximation $\mathbf{\hat R}$.
The output is an RGB value.
We keep the NNs for the diffuse and specular components to be separate and independently process the RGB channels.
The corresponding approximations using NNs are denoted as $\hat{\mathbf D}$ and $\hat{\mathbf S}$, respectively.
The network architecture is shown in \refFig{Architecture}.
The input to the network is a 12-dimensional vector and differs between diffuse and specular NNs.
Both consume the parameters of a single illumination lobe (direction and variance).
However, the diffuse net consumes the normal while the specular net consumes the half-angle.
All layers are full convolutional with 288 units in each layer.
The networks are trained from 200k samples from renderings of spheres produced using Blender. 
An evaluation of this architecture is found in \refFig{NNEvaluation}.

\myfigure{Architecture}{
Our diffuse (\emph{orange}) and specular (\emph{blue}) neural network architecture, that consumes either normal and a single illumination lobe, or half-angle (\emph{left}) and a lobe to produce a color (\emph{right}).
}

\mypara{Prior}
As reflectance is typically more chromatic than illumination is, our prior penalizes the variance of the illumination lobe colors \ie their RGB weights, as in 
$
p(\mathsf\Theta) 
=
\mathrm V(q(\mathsf\Theta))
=
\mathrm E(q(\mathsf\Theta)^2)-
\mathrm E(q(\mathsf\Theta))^2
$, 
where
$
q(\mathsf\Theta)=
\sum_{\Theta\in\mathsf\Theta}
\sum_{i=1}^{n_\mathrm p}
\Theta_{\mathrm w,i}
$.
In other words by, first computing the average color of all lobes $q(\mathsf\Theta)$ and second the variance $\mathrm V(q(\mathsf\Theta))$ of those three channels.

\mypara{Optimization}
Armed with a fast method (see above) to evaluate the cost function, we employ LBFGS~\cite{LBFGS:97} in combination with randomization.
As the full vector $\mathcal O$ does not fit into memory, we use a randomized subset that fits GPU memory in each iteration and dynamically change the randomization set across iterations. We stop our optimization when each observation on average has been sampled 5 times.

\mysubsection{Rendering}{Rendering}
For rendering, the result of the optimization is simply converted into an HDR environment map image by evaluating the estimated PMM for each pixel. Such an environment map along with estimated diffuse/specular parameters are then used with standard offline and online rendering applications as shown in our results.

%% file: results.tex
\mysection{Results}{Results}

\mycfigure{CarResult}{Results on the \lapd\ dataset of four images of police cars with two materials.
The first row shows the input images.
The second row the reflectance maps.
The observed ones are marked with black circles.
In this example, all are observed.
When an RM is not observed, it is re-synthesized.
The third row shows our estimated illumination.
Recall, that it is defined in camera space.
The fourth row contains a re-synthesis using our material and illumination.
Please note, that such a re-synthesis is not possible using a trivial factorization as all images have to share a common material that sufficiently explains the images.
The last row shows a re-synthesis from a novel view, as well as a rendering of the materials in a new (Uffizi) illumination.}

\mycfigure{DockstaResult}{Results on the \docksta\ dataset of six images of IKEA furniture with seven materials in the protocol of \protect\refFig{CarResult}.}

\mycfigure{EamesResult}{Results on \eames\ dataset of six images of a celebrated Eames chair with four materials in the protocol of 
\protect\refFig{CarResult}.
The last row shows a pair of novel views, under input illumination where the first image has unobserved, the second an observed material.}

\myfigure{GaussCoverage}{
    Here, we show the effect of Gauss-sphere coverage: Even for non-round objects with flat areas that have a bad coverage of the Gauss sphere the reconstruction (left) is similar to the reference (right).
}

In this section, we evaluate our approach on synthetic and real image collection data (\refSec{Evaluation}), compare to alternatives (\refSec{Comparison}) and give examples of potential applications (\refSec{Applications}). 
We use L-BFGS solver for all the experiments. The complexity of our optimization in terms of the number of variables is ($7m + 6n_pn$) and hence is linear in terms of the number of input entries in the material$\times$environment observation matrix. For example, for a five-photo, five-material matrix dataset, it costs about 30 minutes using a NVIDIA Titan X GPU. Pre-training the reflection operator $\mathbf R$, both diffuse and specular components, takes about three hours on the same specification.

\mysubsection{Evaluation}{Evaluation}

\mysubsubsection{Datasets}{Datasets}
We evaluated our method using three types of data sets, each holding multiple image collections acquired in different ways. The full resolution images and result images/video are included in the supplementary material.

The first comprises of \synthetic\ images rendered using Mitsuba~\cite{Mitsuba} and a collection of HDR environment maps.
Note that here we have access to ground-truth per-pixel normals and material labels.
Here, we have rendered 3 objects in 3 different scenes with both spheres and real-world shapes that allow synthetic re-combination in an arbitrary fashion.
This allows us to evaluate the proposed approach under different input variants, validating its scalability.

%
The second data set consists of real images collected from the \internet.
We have manually selected the images (using iconic object name-based Google search) and masked the image content.
This dataset has three sets of photographs: the LAPD car (\lapd), the Docksta table (\docksta), and the Eames DSW chair (\eames).
For geometry, we estimated used coarse quality meshes available from ShapeNet. 
Images are good for qualitative evaluation but do not allow to quantify the error, especially in novel views. 

The third dataset contains \photos\ we have taken from designer objects we choose under illumination conditions (in our labs and offices). 
We have 3D-scanned these objects (using Kinect) to acquire their (rough) geometry.
The photos are taken in five different environments and 7 materials are considered.

\mysubsubsection{Qualitative evaluation}{QualitativeResults}

\mypara{Visual quality}
We show results of our approach in Figures \ref{fig:CarResult}, \ref{fig:DockstaResult}, \ref{fig:EamesResult}, and \ref{fig:GowerStreetResult}.
We evaluate the main objective of our approach, \ie estimating illumination and reflectance from a photo collection.  
In each figure, we show the input images, rendering of all objects' materials from original view (with the background from input images) and a novel view as well as visualizations of the material and illumination alone.
Input images are shown on the top with the outputs we produce on the bottom (see supplementary for full images). 
Observed reflectance maps are shown encircled in black.
The objects are rendered from an identical novel view, which is more challenging than rendering from the original view.
The material is shown by re-rendering it under a new illumination.
The exposure between all images is identical, indicating that the reflectance is directly in the right order of magnitude and can transfer to new illuminations.
While the illumination alone does not appear natural, shadow and shading from it produce plausible images, even of unknown materials or new objects. Recall that large parts of the objects are not seen in any of the input images and hence large parts of the environment maps are only estimated from indirect observation. 
Recall that our method does not use any data-driven prior to regularize the results. 

The \lapd\ in \refFig{CarResult} shows a single object made from multiple materials.
\refFig{DockstaResult} shows many chairs in many photos with one common `linking' object \docksta: the chair with material $M_1$, that is used to calibrate all the other objects, which are only observed sometimes.
\refFig{EamesResult} shows multiple Eames chairs from a set of photos \eames.
All show plausible highlights and colors, albeit only observing a fraction of the combinations.
Please see the supplemental video for an animation of the highlights under changing view or object rotations.
\refFig{GowerStreetResult} shows photos taken for this work, with all objects in all illumination conditions.
Note, that both vases are made of multiple materials.
The geometry of all objects in this part (except the chairs) is very approximate and acquired by a depth sensor.
Still a good result quality is possible.

\mypara{Prediction}
Using the \eames\ data sets, we are able to test the predictive ability of our approach by {\em leaving out} one image form the collection and compare it to the acquired ground truth.
This is seen in \refFig{LeaveOneOut}, where starting from the first image, we predict the red chair in the second image and the yellow chair in the third image.

\myfigure{LeaveOneOut}{Estimating materials and illumination using all chairs except the rendered one. Left: reference image; Middle: the red chair is rendered; Right: the yellow chair is rendered.}

\mypara{Progressive estimation}
A key property of our approach is to consolidate information from multiple images to disambiguate material and illumination.
This characteristic implies that adding more images to the photo set should reduce the error.
\refFig{Progressiveness} confirms the rise in performance as more images are added to the linked set.

\myfigure{Progressiveness}{
Progression of quality from left to right.
Every row shows, for a selected material what the additional images can add to the quality in terms of re-rendering, material, and illumination.
}

\mysubsubsection{Quantitative evaluation}{QuantitativeResults}
We evaluate the effect of certain aspects of our method on the end-result (\refFig{MatrixError}).
The error is quantified as DSSIM~\cite{wang2004image} structural image distance (smaller distance indicates better match) between a reference image rendered with known illumination and material compared to another rendering using our estimated material and illumination.
Images were gamma-corrected and linearly tone-mapped before comparison.

\myfigure{MatrixError}{
Effect of different input properties \emph{(horizontal)} on the quality of our approach in terms of the DSSIM error \emph{(vertical, less is better)}: matrix size, structure, and alignment error.}

\mypara{Matrix size}
In \refFig{MatrixError}a, we show the effect of increasing matrix size on the error of predicting the entries for the \synthetic\ data set.
Here, the matrix $\mathcal O$ is complete, \ie all material-illumination pairs are observed.
We see, that with increasing size, the estimation for all entries gets more correct while the task being solved in some sense is also bigger (more different illuminations). Note that the total error residue can go up, but the estimation gets more accurate (compared to the ground truth). 

When the matrix is reduced to a single row or column (\refFig{MatrixError}b) our approach can still estimate illumination and material.
For a $1\times5$ matrix, which estimates a single material form multiple illuminations, the approach does well; but slightly degrades for a $5\times1$ setting, where multiple objects are seen under the same illumination. 

\mypara{Label quality}
We assume the input images to have per-pixel normal and material labels.
In \refFig{MatrixError}d, we study the effect of incorrect normal estimates by adding a different label of uniform noise to the normal. 
We see that good normal fair better with the error in the order of one percent, while larger errors produce an error that saturates still at a low total value. Also, note that for the \internet\ and the \photos\ datasets, the geometry models are coarse and/or noisy. But in absence of ground truth, we could not measure estimation error.

\mysubsection{Comparison}{Comparison}
We compare possible alternatives to our approach as shown in \refFig{Comparison} and supplementary material.
A simple approach could be using image-based rendering based approaches~\cite{shum2008image}, however, these approaches require either flat geometry or a high-enough number of images to reproduce specular appearance, neither of which is available in our input images that show a single image of one condition only.
Effectively, IBR would amount to projective texturing in our setting, that is compared to our approach in  \refFig{Comparison}a.
An alternative could be to run a general intrinsic image approach \cite{Barron2015} on the input images and use the average pixel color of the albedo $k_\mathrm d$ image as the diffuse albedo.
The specular could then be the color that remains.
While this would provide a specular value $k_\mathrm s$, it is not clear how to get a glossiness value $g$ (see \refFig{Comparison}b).

\myfigure{Applications}{Various photo-realistic image manipulations (\eg object insertion)  made possible using our estimated material and environment parameters (see  \protect\refSec{Applications} for details).}

\mysubsection{Application}{Applications}
A typical application of our approach is photo-realistic manipulation of objects in Internet images as shown in \refFig{Applications}.
Having estimated the material and illumination parameters from all the images, we can insert virtual replica into the image (\refFig{Applications}b, \ref{fig:Applications}d),
transfer reflectance estimated from other Internet images to new scenes
(\refFig{Applications}a and \refFig{Applications}c), or introduce new object with material under the estimated illumination. 
Please note that the estimated environment maps were used to render object shadows on (manually added) ground planes (\refFig{Applications}a, \ref{fig:Applications}c, and \ref{fig:Applications}d). 

\mysubsection{User Study}{UserStudy}
We have compared our approach to the similar approach (SIRFS) that extract intrinsic images and lighting in a user study.
When asking $N=250$ subjects if one of five animated turn-table re-renderings using our material information or the model of SIRFS is preferred when showing both in a space-randomized 2AFC the mean preference was $86.5\,\%$ in our favor (std.\ error of the mean is $2.1\,\%$).
The chart of the user response, their mean, the exact sample counts and standard errors for individual images are  presented in \refFig{Study}.

\myfigure{Study}{User study results. the vertical axis is the preference for ours, so more is better. Kinks are standard error of the mean, where small means certainty about the outcome.}

\mycfigure{GowerStreetResult}{Results on the \photos\ of five images of furniture and objects with seven materials in the protocol of 
\protect\refFig{CarResult}.}

\mycfigure{Prior}{
    Effect of increasing prior.
    The top three rows show a 3$\times$3 input.
    The next row shows illumination.
    Following the GT left, we increase the prior weight left to right.
    We note that illumination chromaticity decreases with increasing weight and that a good trade-off is likely at 0.1.
}

\mysubsection{Limitations}{Limitations}
Our approach has three main limitations: 
First, we use a distant light model, which results in estimation errors in large rooms with interior lights. 
Second, although our environment map estimates lead to photo-realistic back projections and predictions, in absence of any data-driven regularization the illumination estimates themselves may look unnatural. This is primarily due to the limited samplings we have in our input measurements. Further, when objects are in close proximity, they may `show up' in environment map estimations. For example, in \refFig{EamesResult}, we see the chair's present red shading. This is because the incoming light for the white chair is distorted by the reflection of the red chair as shown in P4 (please refer to the video in the supplementary material).


%% file: conclusion.tex
\mysection{Conclusion}{Conclusion}
We presented a novel optimization formulation for joint material and illumination estimation from collections of Internet images when different objects are observed in varying illumination conditions sharing coupled material and/or illumination observations. We demonstrated that such a linked material-illumination observation structure can be effectively exploited in a scalable optimization setup to recover robust estimates of material (both diffuse and specular) and effective environment maps. The estimations can then be used for a variety of compelling and photo-realistic image manipulation and object insertion tasks.

This work opens up several interesting future directions to pursue. In this work, we manually curated the photo sets to test our approach. While collecting datasets with star structure for the observation matrix is not so difficult -- one can search with single keywords, especially for iconic designs -- gathering more data linked with tighter connections (\eg loops, full matrix, etc.) is more challenging. One option would be to harvest user annotations and existing links (like in Houzz or Pinterest websites) to collect such data. Such data sets can open up new material-illumination estimation pipelines -- this is particularly exciting as we can update our estimates in an incremental way leading to simultaneous improvement of all the associated estimates. 
Another interesting direction would be to improve the environment map estimates -- one option would be to additionally use data priors to project the estimates environment maps to some data-driven manifold of environment maps measurements. 
Finally, an interesting future direction is to use the material and illumination estimates to improve geometry and pose estimates by refining correspondences by (partially) factoring our shading and illumination effects.